\documentclass[12pt]{iopart}
\usepackage{iopams,epsf,psfig}
\begin{document}
\title[Finite-Dimensional ${\cal PT}$-Symmetric Hamiltonians]{Finite-Dimensional ${\cal PT}$-Symmetric Hamiltonians}

\author{Carl M. Bender, Peter N. Meisinger, and Qinghai Wang}

\address{Department of Physics, Washington University, St. Louis, MO 63130, USA}

\begin{abstract}
This paper investigates finite-dimensional representations of ${\cal
PT}$-symmetric Hamiltonians. In doing so, it clarifies some of the claims made
in earlier papers on ${\cal PT}$-symmetric quantum mechanics. In particular, it
is shown here that there are two ways to extend real symmetric Hamiltonians into
the complex domain: (i) The usual approach is to generalize such Hamiltonians to
include complex Hermitian Hamiltonians. (ii) Alternatively, one can generalize
real symmetric Hamiltonians to include complex ${\cal PT}$-symmetric
Hamiltonians. In
the first approach the spectrum remains real, while in the second approach the
spectrum remains real if the ${\cal PT}$ symmetry is not broken. Both
generalizations give a consistent theory of quantum mechanics, but if $D>2$, a
$D$-dimensional Hermitian matrix Hamiltonian has more arbitrary parameters than
a $D$-dimensional ${\cal PT}$-symmetric matrix Hamiltonian.
\end{abstract}

\submitto{\JPA}

It has been observed that non-Hermitian Hamiltonians that exhibit ${\cal PT}$
symmetry can have real spectra. For example, the class of non-Hermitian
Hamiltonians
\begin{eqnarray}
H=p^2+x^2(ix)^\nu
\label{eq1}
\end{eqnarray}
have positive real discrete spectra so long as $\nu>0$ and appropriate
boundary conditions are specified \cite{r1,r2,r3}. The
domain $\nu>0$ is the region of {\em unbroken} ${\cal PT}$ symmetry, while $\nu
<0$ is the region of {\em broken} ${\cal PT}$ symmetry. The distinction between
these two regions is as follows: When $\nu>0$, the eigenstates of $H$ are also
eigenstates of ${\cal PT}$, but when $\nu<0$, the eigenstates of $H$ are not
eigenstates of ${\cal PT}$. In the unbroken region the eigenvalues of $H$ are
real and in the broken region some eigenvalues of $H$ may be real, but the rest
appear as complex-conjugate pairs.

In a recent letter it was shown that in the region of unbroken ${\cal PT}$
symmetry a ${\cal PT}$-symmetric Hamiltonian possesses an additional symmetry
represented by the complex linear operator ${\cal C}$ \cite{r4}. The operator
${\cal C}$ commutes with $H$ and with ${\cal PT}$ and can be used to construct
an inner product whose associated norm is positive. The theory defined by the
complex Hamiltonian (\ref{eq1}) with $\nu>0$ is a fully consistent and unitary
theory of quantum mechanics \cite{r4}.

One might conjecture that ${\cal PT}$ symmetry is a generalization of
Hermiticity. However, as we will argue in this paper, this view is not quite
precise. Rather, we will argue that the appropriate way to construct complex
Hamiltonians is to begin with a real symmetric Hamiltonian and to extend the
matrix elements into the complex domain in such a way that certain constraints
are satisfied. There are two distinct ways to perform this construction. First,
one can generalize real symmetric Hamiltonians to the case of Hermitian
Hamiltonians and second, one can generalize real symmetric Hamiltonians to the
case of ${\cal PT}$-symmetric Hamiltonians that are not Hermitian. In the second
generalization the symmetry of the Hamiltonian is maintained but the matrix
elements are allowed to become complex with the condition that the ${\cal PT}$
operator commutes with $H$.

Many of the Hermitian Hamiltonians commonly studied in quantum mechanics are
actually real and symmetric. For example, this is the case of the Hamiltonian
representing a particle in a real potential $V(x)$, so that $H=p^2+V(x)$; this
Hamiltonian is explicitly real \cite{PAR}. To show that it is symmetric we
display it as a continuous matrix in coordinate space:
\begin{eqnarray}
H(x,y)=-{d\over dx}{d\over dy}\delta(x-y)+V\left({x+y\over2}\right)\delta(x-y).
\label{eq2}
\end{eqnarray}
This matrix is explicitly symmetric under the interchange of $x$ and $y$. The
${\cal PT}$-symmetric Hamiltonian in (\ref{eq1}) is also symmetric in
coordinate space; however, it is complex for all
$\nu>0$.\footnote{This Hamiltonian is complex even when $\nu$ is a positive
even integer because the boundary conditions associated with the eigenvalue
problem $H\phi=E\phi$ are complex. See Ref.~\cite{r1}.}

\begin{figure}[b!]
\vspace{3.3in}
\includegraphics{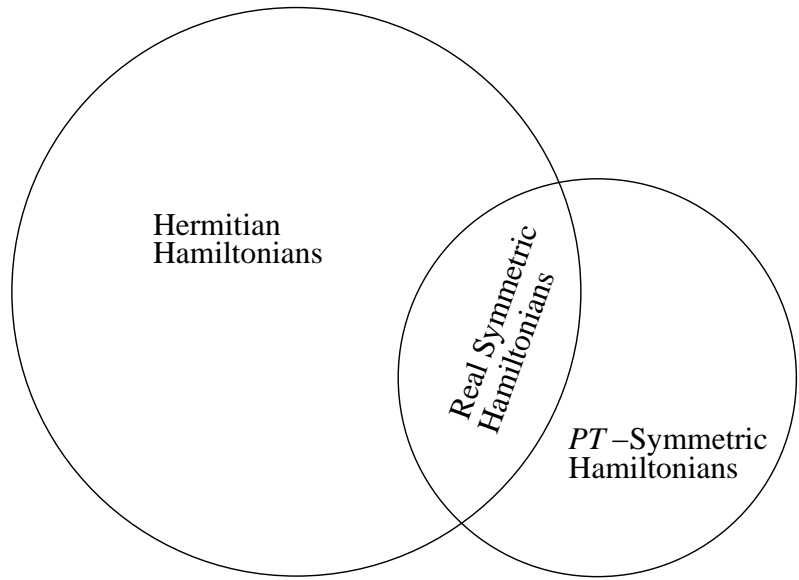}
\caption{Venn diagram showing that the intersection between the classes of
Hermitian and ${\cal PT}$-symmetric matrix Hamiltonians is the class of real
symmetric matrix Hamiltonians.}
\label{f1}
\end{figure}

In this paper we investigate the case of finite-dimensional matrix Hamiltonians.
We show that Hermitian matrix Hamiltonians and ${\cal PT}$-symmetric matrix
Hamiltonians are both acceptable generalizations of real symmetric matrix
Hamiltonians. Furthermore, they define consistent theories of quantum mechanics.
We also demonstrate that for the case of $D$-dimensional matrices
the class of Hermitian matrix Hamiltonians is much larger than the class of
${\cal PT}$-symmetric matrix Hamiltonians. Specifically, we know that for large
$D$ the number of real parameters in a real symmetric matrix is asymptotically
${1\over2}D^2$ and the number of real parameters in a Hermitian matrix is $D^2$.
We will see that the number of real parameters in a ${\cal PT}$-symmetric matrix
Hamiltonian is asymptotically ${3\over4}D^2$. The overlap between the classes of
Hermitian and ${\cal PT}$-symmetric matrix Hamiltonians is only the class of
real symmetric matrices. A Venn diagram showing the relationships between the
classes of Hermitian, ${\cal PT}$-symmetric, and real symmetric matrix
Hamiltonians is given in Fig.~\ref{f1}.

To construct a finite-dimensional ${\cal PT}$-symmetric matrix Hamiltonian we
begin by defining the operators that represent time reversal ${\cal T}$ and
parity ${\cal P}$. Both of these operators represent discrete reflection
symmetries and thus we must have ${\cal T}^2 = {\cal P}^2=1$. Furthermore, we
assume ${\cal T}$ and ${\cal P}$ are independent operators, so that they commute
$[{\cal P},{\cal T}]=0$. For simplicity we define the time reversal
operator as complex conjugation. One can also define ${\cal T}$ to be Hermitian
conjugation (complex conjugation and transpose). However, we will see that
because all of the relevant matrices in the theory are symmetric it makes no
difference whether ${\cal T}$ performs a transpose. It is also possible to
choose a more complicated definition for ${\cal T}$. For example, ${\cal T}$
could be the combined action of complex conjugation and multiplication by some
complex matrix. Such alternative definitions for ${\cal T}$ will be considered
later and we will argue that without any loss of generality we may define
${\cal T}$ to be just complex conjugation.

Next, we consider the operator ${\cal P}$ representing parity. Since ${\cal P}$
commutes with ${\cal T}$, the entries in the matrix representing ${\cal P}$ are
all real. Furthermore, we will see that ${\cal P}$ must be symmetric. (If it is
not symmetric, then we will find that the ${\cal C}$ matrix that we will
ultimately construct will not commute with the Hamiltonian $H$. As a result,
the quantum theory will violate unitarity. We will return to this point later
on.)

The fact that ${\cal P}^2=1$ implies that all the eigenvalues of ${\cal P}$ are
either $+1$ or $-1$. To construct the most general $D$-dimensional matrix ${\cal
P}$ let us suppose that there are $m_+$ positive eigenvalues and $m_-$ negative
eigenvalues, where $m_++m_-=D$. That is, if ${\cal P}$ is diagonalized, then it
has the form
\begin{eqnarray}
{\cal P}_0={\rm diag}\{1,1,\cdots,1,-1,-1,\cdots,-1\}.
\label{eq3}
\end{eqnarray}
The most general parity matrix can be expressed as
\begin{eqnarray}
{\cal P}=R{\cal P}_0R^{-1},
\label{eq4}
\end{eqnarray}
where $R$ is the most general $D$-dimensional rotation (orthogonal) matrix
\cite{r5}.\footnote{Of course, one could take the matrix $R$ to be more general
than orthogonal by choosing it to be unitary. However, in this case the parity
operator ${\cal P}$ will be complex and will not commute with ${\cal T}$.}

There are ${1\over2}D(D-1)$ arbitrary parameters in the matrix $R$. However,
there are fewer than this number of parameters in the matrix ${\cal P}$ in
(\ref{eq4}). Indeed, if $m_-=0$ so that ${\cal P}_0$ is the identity matrix,
then there are no arbitrary parameters in ${\cal P}$. The exact number of
arbitrary parameters in ${\cal P}$ is given by the formula
\begin{eqnarray}
{1\over2}D(D-1)- {1\over2}m_+(m_+-1)-{1\over2}m_-(m_--1).
\label{eq5}
\end{eqnarray}
Clearly, when $D$ is even, ${\cal P}$ has the greatest number of arbitrary
parameters if $m_+=m_-={1\over2}D$. When $D$ is odd, the number of parameters
is maximized if we choose $m_+-m_-=1$;
that is, $m_+={1\over2}(D+1)$ and $m_-={1\over2}(D-1)$. Thus, for all $D$, the
greatest number of parameters in ${\cal P}$ is given by the formula
\begin{eqnarray}
{1\over4}D^2-{1\over8}\left[1-(-1)^D\right].
\label{eq6}
\end{eqnarray}

Let us illustrate these results. The most general one-dimensional parity matrix
${\cal P}=1$ has no free parameters. The most general two-dimensional parity
matrix has one parameter:
\begin{eqnarray}
{\cal P}=\left(
\begin{array}{cc}
\cos\phi&\sin\phi\\\sin\phi&-\cos\phi\\
\end{array}
\right).
\label{eq7}
\end{eqnarray}
The most general three-dimensional parity matrix has two parameters:
\begin{eqnarray}
{\cal P}=\left(
\begin{array}{ccc}
\cos^2\phi-\sin^2\phi\,\cos2\theta&\sin2\phi\,\cos\theta&-\sin^2\phi\,\sin2\theta\\
\sin2\phi\,\cos\theta&-\cos2\phi&\sin2\phi\,\sin\theta\\
-\sin^2\phi\,\sin2\theta&\sin2\phi\,\sin\theta&\cos^2\phi+\sin^2\phi\,\cos2\theta\\
\end{array}
\right).
\label{eq8}
\end{eqnarray}

Now, let us construct the most general $D$-dimensional ${\cal PT}$-symmetric
matrix Hamiltonian $H$, where by ${\cal PT}$-symmetric we mean that the operator
${\cal PT}$ commutes with $H$. We will assume that the matrix $H$ is symmetric.
(If $H$ were not symmetric, then the eigenvectors of $H$ would not be
orthogonal.\footnote{The inner product here is just the ordinary dot product,
${\bf v}\cdot{\bf v}\equiv {\bf v}^{\rm T}{\bf v}$.} We will consider the
possibility of an asymmetric $H$ later.) To count the number of parameters in
$H$ we take the parity matrix to be in diagonal form ${\cal P}_0$ as in
(\ref{eq3}). If the operator ${\cal P}_0{\cal T}$ commutes with $H_0$,
\begin{eqnarray}
{\cal P}_0 H_0^*= H_0{\cal P}_0,
\label{eq9}
\end{eqnarray}
then $H_0$ has the $2\times2$ block form 
\begin{eqnarray}
H_0=\left(
\begin{array}{cc}
A&iB\\iB^{\rm T}&C\\
\end{array}
\right),
\label{eq10}
\end{eqnarray}
where $A$ is a real symmetric $m_+\times m_+$ matrix, $C$ is a real symmetric
$m_-\times m_-$ matrix, and $B$ is a real $m_+\times m_-$ matrix. Thus, the
number of parameters in $H_0$ is ${1\over2}D(D+1)$. We then transform ${\cal
P}_0$ to ${\cal P}$ using the rotation matrix $R$, and find that the number of
arbitrary real parameters in the corresponding ${\cal PT}$-symmetric Hamiltonian
$H=RH_0R^{-1}$ is given by the combined number ${1\over2}D(D+1)$ of free
parameters in $H_0$ and the number (\ref{eq6}) of free parameters in ${\cal P}$.
Since $H_0$ is not Hermitian (it is complex and symmetric) and $R$ is
orthogonal, as we have argued above, it follows that the Hamiltonian $H$ is
non-Hermitian and is not unitarily equivalent to any Hermitian matrix.

As an example, for the case $D=2$ the most general ${\cal PT}$-symmetric
Hamiltonian, where ${\cal P}$ is given in (\ref{eq7}), contains four free
parameters and has the form
\begin{eqnarray}
H=\left(
\begin{array}{cc}
r+t\cos\phi-is\sin\phi&is\cos\phi+t\sin\phi\\is\cos\phi+t\sin\phi&r-t\cos\phi
+is\sin\phi\\
\end{array}
\right).
\label{eq11}
\end{eqnarray}
The most general $3\times3$ ${\cal PT}$-symmetric Hamiltonian has eight free
parameters.

For arbitrary $D$ there are 
\begin{eqnarray}
{3\over4}D^2+{1\over2}D-{1\over8}\left[1-(-1)^D\right]
\label{eq12}
\end{eqnarray}
real parameters in the most general ${\cal PT}$-symmetric matrix Hamiltonian.
For purposes of comparison, in Table \ref{t1} we give formulas for the number of
free parameters in the most general real symmetric $D\times D$ parity matrix,
the most general matrix $H_0$ that commutes with ${\cal P}_0{\cal T}$, the most
general ${\cal PT}$-symmetric matrix $H$, the most general Hermitian matrix
Hamiltonian, and finally the most general real symmetric matrix Hamiltonian.

\begin{table}
\caption[t1]{Number of arbitrary real parameters in the following most general
$D\times D$ matrices: (i) real symmetric parity ${\cal P}$, (ii) ${\cal
P}_0{\cal T}$-symmetric $H_0$, (iii) ${\cal PT}$-symmetric $H$, (iv) Hermitian
$H$, and (v) real symmetric $H$.}
\begin{center}
\begin{tabular}{|c||c|c|c|c|c|c||c|}
\hline\hline
Dimension $D$&1&2&3&4&5&6&Large $D$\\
\hline\hline
Real Symmetric ${\cal P}$: ${1\over4}D^2-{1\over8}\left[1-(-1)^D\right]$&0&1&2&4&6&9&$\sim {1\over4}D^2$\\
\hline
${\cal P}_0{\cal T}$-Symmetric $H_0$: ${1\over2}D(D+1)$&1&3&6&10&15&21&$
\sim{1\over2}D^2$\\
\hline
${\cal PT}$-symmetric $H$:
${3\over4}D^2+{1\over2}D-{1\over8}\left[1-(-1)^D\right]$&1&4&8&14&21&30&$\sim{3\over4}D^2$\\
\hline\hline
Hermitian $H$: $D^2$ &1&4&9&16&25&36&$D^2$\\
\hline\hline
Real Symmetric $H$: ${1\over2}D(D+1)$&1&3&6&10&15&21&$\sim{1\over2}D^2$\\
\hline\hline
\end{tabular}
\medskip
\label{t1}
\end{center}
\end{table}

Once we have found the most general ${\cal PT}$-symmetric matrix Hamiltonian 
we proceed according to the recipe described in Ref.~\cite{r4}. First, we
find the energy eigenvalues. The eigenvalues for $H$ in (\ref{eq11}) are
\begin{eqnarray}
\varepsilon_\pm = r\pm t\cos\alpha,
\label{eq13}
\end{eqnarray}
where $\sin\alpha=s/t$ and the unbroken ${\cal PT}$-symmetric region is
$s^2\leq t^2$. 

Next, we find the corresponding eigenstates:
\begin{eqnarray}
|\varepsilon_\pm)={1\over\sqrt{2(1\mp\cos\alpha)\cos\alpha}}
\left( \begin{array}{c}
\sin\alpha \cos{\phi\over2}-i(1\mp \cos\alpha)\sin{\phi\over2}\\
\sin\alpha \sin{\phi\over2}+i(1\mp \cos\alpha)\cos{\phi\over2}\\
\end{array}
\right).
\label{eq14}
\end{eqnarray}
Because we are in the {\it unbroken} ${\cal PT}$-symmetric region these states
are also eigenstates of the ${\cal PT}$ operator. We have chosen the phase in
(\ref{eq14}) so that the eigenvalue under the ${\cal PT}$ operator is unity:
\begin{eqnarray}
{\cal PT}|\varepsilon_+)&=&|\varepsilon_+),\nonumber\\
{\cal PT}|\varepsilon_-)&=&|\varepsilon_-).
\label{eq17}
\end{eqnarray}

It seems appropriate now to define an inner product with respect to the
${\cal PT}$ operator. To do so we define the ${\cal PT}$ conjugate $(\cdot|$
of the state $|\cdot)$ as follows:
\begin{eqnarray}
(\cdot| \equiv \left[ {\cal PT}|\cdot)\right]^{\rm T},
\label{eq101}
\end{eqnarray}
where ${\rm T}$ is matrix transpose. The ${\cal PT}$ inner product of two states
$|a)$ and $|b)$ is now defined as the dot product of the ${\cal PT}$ conjugate
of $|a)$ and $|b)$:
\begin{eqnarray}
(a|b) \equiv \left[ {\cal PT}|a)\right]^{\rm T} \cdot |b).
\label{eq102}
\end{eqnarray}
This inner product has the symmetry property $(a|b)^*=(b|a)$.

By virtue of (\ref{eq17}), for the eigenstates of the Hamiltonian the state
$(\varepsilon_\pm|$ is just the transpose of $|\varepsilon_\pm)$. The states in
(\ref{eq17}) are normalized so that their ${\cal PT}$ norms are
\begin{eqnarray}
(\varepsilon_+|\varepsilon_+)&=&1,\nonumber\\
(\varepsilon_-|\varepsilon_-)&=&-1.
\label{eq16}
\end{eqnarray}
Also, the matrix Hamiltonian is symmetric, so these states are orthogonal
with respect to the ${\cal PT}$ inner product:
\begin{eqnarray}
(\varepsilon_+|\varepsilon_-)=(\varepsilon_-|\varepsilon_+)=0.
\label{eq15}
\end{eqnarray}

Finally, we construct the ${\cal C}$ operator as outlined in Ref.~\cite{r4}:
\begin{eqnarray}
{\cal C}&=&|\varepsilon_+)(\varepsilon_+|+|\varepsilon_-)(\varepsilon_-|
\nonumber\\
&=& {1\over\cos\alpha}\left(
\begin{array}{cc}
\cos\phi-i\sin\alpha\sin\phi& \sin\phi+i\sin\alpha\cos\phi\\
\sin\phi+i\sin\alpha\cos\phi& -\cos\phi+i\sin\alpha\sin\phi\\
\end{array}
\right).
\label{eq18}
\end{eqnarray}
It is easy to verify that the matrix ${\cal C}$ commutes with ${\cal PT}$
and with $H$ and that ${\cal C}^2=1$. The eigenstates of the Hamiltonian
are simultaneously eigenstates of ${\cal C}$:
\begin{eqnarray}
{\cal C}|\varepsilon_+)&=&+|\varepsilon_+),\nonumber\\
{\cal C}|\varepsilon_-)&=&-|\varepsilon_-).
\label{eq19}
\end{eqnarray}

Using these results we can define a {\it new} inner product in which the {\it
bra} states are the ${\cal CPT}$ conjugates of the {\it ket} states:
\begin{eqnarray}
\langle\cdot|\equiv\left[{\cal CPT}|\cdot\rangle\right]^{\rm T}.
\label{eq103}
\end{eqnarray}
The ${\cal CPT}$ inner product of two states $|a\rangle$ and $|b\rangle$ is now
defined as the dot product of the ${\cal CPT}$ conjugate of $|a\rangle$ and
$|b\rangle$:
\begin{eqnarray}
\langle a|b\rangle \equiv \left[{\cal CPT}|a\rangle\right]^{\rm T} \cdot
|b\rangle.
\label{eq104}
\end{eqnarray}
This inner product has the symmetry property $\langle a|b\rangle^*=\langle b|a
\rangle$. The advantage of the ${\cal CPT}$ inner product is that the
associated norm is positive definite.

We recover the parity operator 
\begin{eqnarray}
{\cal P}=\left(
\begin{array}{cc}
0&1\\1&0\\
\end{array}
\right)
\label{eq20}
\end{eqnarray}
that was used in Ref.~\cite{r4} by choosing $\phi=\pi/2$. All the results that
are reported in Ref.~\cite{r4} are also obtained for this choice of $\phi$.
However, there is an error in Ref.~\cite{r4}. In this reference the parameters
$s$ and $t$ in the Hamiltonian must be identical; they cannot be unequal because
then the matrix would not be symmetric and the eigenvectors would not be
orthogonal.

What happens if we choose the parity operator ${\cal P}$ to have an irregular
distribution of positive and negative eigenvalues? For example, suppose we take
$D=8$ and choose $m_+=6$ and $m_-=2$. [Of course, in this case there are only 
twelve real parameters in ${\cal P}$ instead of the sixteen parameters that
occur in the symmetric case for which $m_+=m_-=4$. See (\ref{eq6}).
Correspondingly, there are also four fewer parameters in the Hamiltonian.]
We have found that the signs of the ${\cal PT}$ norms [see (\ref{eq102})] of the
eigenstates of the Hamiltonian are exactly the same as the signs of the
eigenvalues of ${\cal P}$. However, the {\it order} of the signs depends on the
values of the parameters in $H$ and is unpredictable. The operator ${\cal C}$ is
exactly what is needed to cancel each of the minus signs in the ${\cal PT}$
norm so that the ${\cal CPT}$ norms of the eigenstates are all positive. 

The natural question that arises is whether it is possible to have a more
general formalism for ${\cal PT}$-symmetric matrix Hamiltonians; that is, to
have matrix Hamiltonians with more arbitrary parameters than the number given in
(\ref{eq12}). There are two possibilities: First, one could consider having
an asymmetric matrix Hamiltonian $H$ or an asymmetric parity matrix ${\cal P}$.
Second, we could generalize the time reversal operator to include a matrix
multiplying the complex conjugation operator.

If the matrix Hamiltonian $H$ is not symmetric, then eigenstates of $H$
corresponding to different eigenvalues will not be orthogonal. This forces us to
generalize the ${\cal PT}$ inner product $(\cdot|\cdot)$ to include a {\it
weight matrix} $W$ \cite{rX}. That is, rather than having an ordinary dot
product of vectors, we would have to generalize the definition of the inner
product to $(\cdot|W|\cdot)$, where the matrix elements of $W$ are chosen so
that 
\begin{eqnarray}
(\varepsilon_m|W|\varepsilon_n)=\delta_{mn}.
\label{eq21}
\end{eqnarray}
In this case, the matrix $W$ plays the same role as the operator ${\cal C}$.
The drawback of this generalization is that $W$ will not commute with $H$. As
we now argue, we must reject this generalization of the Hamiltonian because the
theory is not unitary: Unitarity means that the inner product of two states is
independent of time. In the Schr\"odinger picture the states $|a,0)$ and $|b,0)$
at time $t=0$ evolve into the states $|a,t)=e^{-iHt}|a,0)$ and $|b,t)=e^{-iHt}|
b,0)$ at time $t$. Thus,
$$(a,t|=[{\cal PT}|a,t)]^{\rm T}=(a,0|e^{iHt}.$$
The inner product between these states will not be independent of time unless
$e^{iHt}We^{-iHt}=W$ (remember that $H$ commutes with ${\cal PT}$), and this
requires that $W$ and $H$ commute. If $W$ and $H$ do not commute, the theory
must be abandoned because it violates unitarity and is therefore physically
unacceptable.

Similarly, if we generalize the parity operator to the case of an asymmetric
matrix ${\cal P}$, the most general ${\cal PT}$-symmetric $H$ will be
asymmetric. Again, we must reject this possibility.

Finally, we ask if it is possible to generalize ${\cal T}$ so that it is a
product of some matrix $B$ and the complex conjugation operator. The condition
that ${\cal T}^2=1$ implies that $BB^*=1$. Also, the requirement that $[{\cal
P},{\cal T}]=0$ imposes the constraint $[{\cal P},B]=0$. These two conditions
are so strong that no additional parameters appear in the most general ${\cal
PT}$-symmetric matrix Hamiltonian $H$.

We do not believe, as has been claimed (see, for example, Ref.~\cite{rZ} and
references therein), that Hermiticity is a special case of ${\cal PT}$ symmetry.
The problem with the analysis in Ref.~\cite{rZ} is that the norm associated
with the inner product is not positive. To observe this nonpositivity we
construct a vector that is a linear combination of eigenvectors of the
Hamiltonian: $\mu|\varepsilon_m)+\nu|\varepsilon_n)$, where $\mu$ and $\nu$ are 
complex numbers. According to Eq.~(15) in Ref.~\cite{rZ}, the norm of this
vector is $\mu^2+\nu^2$, which is not positive in general.

In conclusion, the matrix constructions presented in this paper have changed our
views regarding the relationship between Hermiticity and ${\cal PT}$ symmetry.
We have found that ${\cal PT}$-symmetric Hamiltonians should not be regarded
as generalizations of Hermitian Hamiltonians; rather, based on our study of
finite matrices we understand that these are two totally distinct and unitarily
inequivalent complex classes of Hamiltonians whose overlap is restricted to the
class of real symmetric Hamiltonians. We conjecture that the picture in
Fig.~\ref{f1} continues to be valid even for infinite-dimensional
coordinate-space Hamiltonians.

\vspace{0.5cm}
\begin{footnotesize}
This work was supported in part by the U.S.~Department of Energy.
\end{footnotesize}

\end{document}